\documentclass[%
 aip,
 amsmath,amssymb,
 reprint,%
]{revtex4-1}

\usepackage{graphicx}
\usepackage{dcolumn}
\usepackage{bm}

\usepackage[utf8]{inputenc}
\usepackage[T1]{fontenc}
\usepackage{mathptmx}
\usepackage{etoolbox}

\makeatletter
\def\@email#1#2{%
 \endgroup
 \patchcmd{\titleblock@produce}
  {\frontmatter@RRAPformat}
  {\frontmatter@RRAPformat{\produce@RRAP{*#1\href{mailto:#2}{#2}}}\frontmatter@RRAPformat}
  {}{}
}%
\makeatother
\begin{document}


\title[]{Impact of Atomic Substitution on Core-Hole Relaxation Dynamics: A Study of Br$_2$ and IBr}
\author{Nivedita Bhat\textsuperscript{\dag}}
\author{Yeonsig Nam\textsuperscript{\dag,*}}%
 \email{nam@anl.gov}
\affiliation{ 
Chemical Sciences and Engineering Division, Argonne National Laboratory, Lemont, Illinois 60439
}%
\author{Linda Young} 
\affiliation{ 
Chemical Sciences and Engineering Division, Argonne National Laboratory, Lemont, Illinois 60439
}%
\affiliation{%
The James Franck Institute and Department of Physics, The University of Chicago, Chicago, Illinois 60637}%
\author{Stephen H. Southworth\textsuperscript{*}}
\email{southworth@anl.gov}
\affiliation{ 
Chemical Sciences and Engineering Division, Argonne National Laboratory, Lemont, Illinois 60439
}%

\author{Phay J. Ho\textsuperscript{*}}
\email{pho@anl.gov}
\affiliation{ 
Chemical Sciences and Engineering Division, Argonne National Laboratory, Lemont, Illinois 60439
}%
\footnotetext{\textsuperscript{\dag} These authors contributed equally}

\date{\today}

\begin{abstract}
Understanding inner-shell decay processes in heavy-element molecules is essential for unraveling x-ray-induced photodynamics and advancing molecular imaging techniques.
In this study, we investigate the influence of atomic substitution on core-hole relaxation dynamics and molecular fragmentation in Br$_2$ and IBr, initiated by x-ray absorption at the Br $K$-edge. 
Using a combination of X-ray/ion coincidence measurements and Monte Carlo/molecular dynamics simulations, we track charge distribution and the kinetic energy release (KER) of fragment ions with a total charge from 2+ to 8+. For both molecules, the simulated KER values show good agreement with experiment across different fragmentation channels.  Our comparison reveals that substituting Br with the heavier I atom in IBr has minimal impact on the inner-shell electronic decay process, but significantly influences nuclear motion,
leading to slower dissociation, thereby a KER close to the Coulomb limit—an effect attributed to the atomic mass.
These findings highlight the interplay between electronic and nuclear effects in molecular fragmentation, particularly in heavy-element species, and provide new insights into medical therapies, structural biology, and astrophysics.
\end{abstract}

\maketitle

\section{INTRODUCTION\label{intro}}
Understanding x-ray–induced processes in molecular environments has a wide range of applications, spanning medical therapies, structural biology, and astrophysics. Radionuclides containing heavy elements such as iodine and bromine have been explored as sources of low-energy ($<$ 1 keV) Auger electrons for targeted cancer treatment \cite{Howell02012023,Ku2019}. However, the generation and spectra of these Auger electrons, which are highly sensitive to the surrounding molecular environment, remain poorly characterized and are often based on atomic data \cite{falzone2017absorbed}. 
Radiation damage, including x-ray–induced structural distortions, also limits the effectiveness of x-ray scattering and diffraction techniques for structural determination of biological matter and 
macromolecular crystals \cite{doi:10.1021/acsnano.7b03447,Stern:dz5149}. In astrophysical environments, stellar x-rays catalyze early dust evolution in protoplanetary disks and drive chemical reactions that influence the formation of complex molecules in the interstellar medium \cite{lepp1996x, gavilan2022laboratory,ciaravella2020x}.

X-rays induce multistep processes in molecules, beginning with atomic site-specific ionization of inner-shell electrons and the creation of core-hole states. In atoms and molecules,
these vacancies decay predominantly via a sequence of radiative (fluorescence) and nonradiative (Auger decay) transitions \cite{Rudek2012,Dunford12:pra}, with the latter incrementally accumulating positive charges.  
Electron transfer redistributes this charge across atomic sites, enhancing electrostatic repulsion and ultimately driving fragmentation via Coulomb explosion, which produces energetic atomic ions.

In synchrotron-based experiments, where inner-shell decay cascades are initiated by single-photon absorption, the charging process is relatively slow (tens of femtoseconds), leaving a window in which nuclear motion can proceed concurrently with electronic decay \cite{Ho23:jcp}. 
By contrast, intense x-ray free-electron laser (XFEL) pulses can promote absorption of multiple photons within a single pulse, creating multiple core holes and accelerating the charging of individual atomic sites to even shorter timescales \cite{rudenko2017femtosecond}. 
These ultrafast charging dynamics provide an alternative implementation of the Coulomb explosion imaging (CEI) technique to probe static and time-dependent molecular structures \cite{doi:10.1126/science.1253607,PhysRevLett.132.123201,Boll2022}. CEI was first demonstrated with swift molecular ions passing through thin solid films, where electron stripping occurred essentially instantaneously ($\sim$100 as, far shorter than nuclear motions) \cite{doi:10.1126/science.244.4903.426}, and was later extended to strong-field optical laser and XFEL pulses with a pulse duration of $\sim$10 to 25 fs\cite{PhysRevLett.132.123201,Boll2022}. Unlike the thin-film case, XFEL and laser pulses act on timescales comparable to nuclear motion, allowing structural rearrangements during the charging process.

Heavy-element molecules, such as bromine (Br$_2$) and iodine bromide (IBr) when excited by single-photon absorption at a synchrotron, are good test systems for studying the effect of atomic substitution (Br to I) on molecular fragmentation dynamics during multi-step core-hole relaxation cascades. 
Polyatomic molecules such as CH$_2$I$_2$ \cite{XLi2025} exhibit intricate fragmentation dynamics, and intense XFEL pulses drive complex charging dynamics owing to pulse-parameter-sensitive multiphoton absorption and thereby creation of multiple core holes during the decay cascade \cite{young2010femtosecond,PhysRevLett.110.173005,Rudek2012,rudenko2017femtosecond, Ho15:pra,Li2021PRL,Erk2013PRL}. 
Within this context, our choice of a linear diatomic molecule studied under synchrotron radiation reduces the complexities arising from multiphoton absorption and multiple fragmentation pathways, and disentangles the role of atomic substitution in electronic cascades and the slower nuclear motion coming from higher mass of I.  

Previous work used x-ray/ion coincidence spectroscopy to compare inner-shell cascades in IBr, initiated by either an iodine or bromine 2p hole\cite{Ho23:jcp}.  To capture molecular effects, a quantum mechanical description of both nuclei and electrons, focusing on the decay of [IBr]$^{1+}$ to [IBr]$^{3+}$ via two Auger events, was employed \cite{doi:10.1021/acs.jctc.4c00778}. In this framework, intramolecular charge redistribution arises naturally from the delocalization and time evolution of the electronic wave function (molecular orbital), rather than from an ad hoc atom-to-atom charge-transfer model.
However, the extent to which atomic substitution (e.g., replacing bromine with iodine) influences charge redistribution and molecular fragmentation processes remains less understood.

\begin{figure*}[h]
\includegraphics[width=14cm]{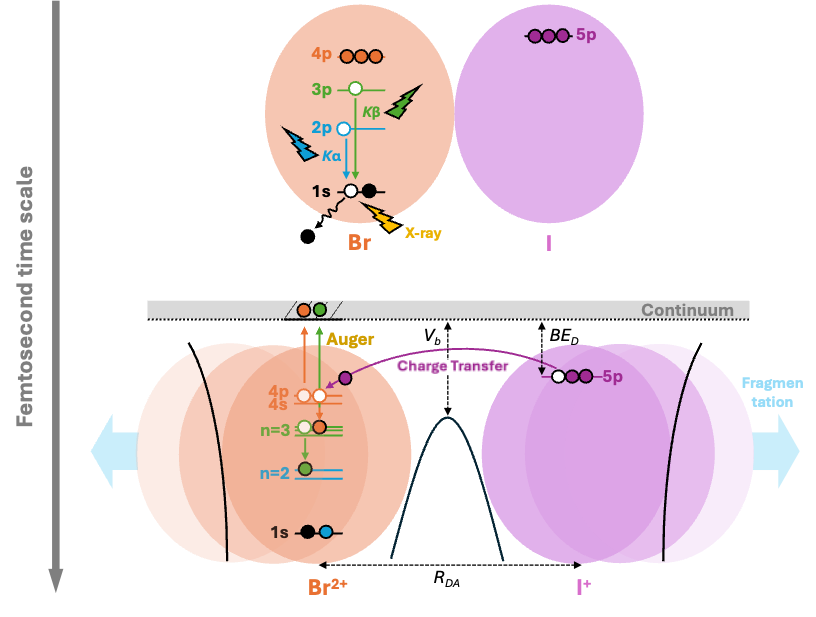}
\caption{Top panel: Neutral IBr (representative; Br$_2$ behaves analogously) exposed to synchrotron X‑rays. A Br 1s electron (black circle) is photoionized, followed by X‑ray fluorescence: either $K\alpha$ (2p $\rightarrow$ 1s) or $K\beta$ (3p $\rightarrow$ 1s), emitting a photon. Filled circles denote electrons; empty circles denote holes; lightning symbols denote photons. 
Bottom panel: Subsequent core-hole relaxation proceeds via successive Auger decays that emit electrons. Electron transfer can occur between electron donor (D) and acceptor (A) at the outermost valence shells (Br: 4p, I: 5p), when the electron binding energy ($BE_D$) is higher than the Coulombic barrier ($V_b$) at internuclear distance ($R_{DA}$); see eq~\ref{Eq1}. The resulting positive charges drive molecular fragmentation via Coulomb repulsion.}
\label{sch1}
\end{figure*}

In this work, we study the impact of atomic substitution on core-hole relaxation dynamics and molecular fragmentation in Br$_2$ and IBr.
We create a Br 1s core hole by exciting at the Br K-edge (see Figure~\ref{sch1}), and tag the ensuing relaxation via creation of 2p or 3p holes by detecting K$\alpha$ and K$\beta$ fluorescence in coincidence with fragment ions, which yields x-ray emission energies and fragment kinetic energy release (KER).
We track the redistribution of charge and the resulting nuclear motion of fragment ions, during core-hole relaxation decay, by Monte Carlo/Molecular Dynamics (MC/MD) simulations\cite{Ho2017:jpb} and a classical over-the-barrier (COB) model\cite{XLi2025}.  Specifically, we focus on the KER distributions and their deviations from Coulomb explosion predictions, highlighting the role of electronic and nuclear effects in shaping the dissociation dynamics.
We show that 
substituting Br with the heavier I atom in IBr has minimal impact on the inner-shell electronic decay process, but its heavier atomic mass significantly influences nuclear motion, leading to slower dissociation and thus a KER closer to the Coulomb limit. 

\section{Methods}
\subsection{EXPERIMENTAL METHODS\label{expt}}
The experiments were conducted on beamline 7-ID at the Advanced Photon Source \cite{Walko16:cp} using the x-ray/ion coincidence spectrometer described in Ref.~\citenum{Ho23:jcp}.
An effusive jet of IBr and Br$_2$ was intersected by monochromatized x-rays tuned above the Br 1s ionization energy of 13482.1(3) eV~\cite{Boudjemia20:pccp}. Br 1s-2p ($K\alpha$) or 1s-3p ($K\beta$) fluorescence, detected by a SiLi spectrometer, triggered ion time-of-flight measurements of coincident I and Br fragment ions. The energy resolution allowed clear separation of decays from Br 2p and 3p vacancies. Further details of the instrumentation are provided in Refs~\cite{Ho23:jcp,Walko16:cp}, and Supporting Information. 
 
The Br 1s vacancies have fairly large total fluorescence yields of 0.618~\cite{Krause79A:jpcrd}, but the $K\beta$/$K\alpha$ ratio $\sim$0.14~\cite{Kortright09:xraydata}. The photoion-photoion coincidence (PIPICO) measurements, made in coincidence with $K\beta$ x-rays, are therefore less intense and more challenging to analyze compared with the $K\alpha$ measurements reported in Ref.~\cite{Ho23:jcp}. An example from the I$^{2+}$--$^{79}$Br$^{2+}$ ion pair is plotted in Fig. S1.  For the IBr and Br$_2$ ion fragmentation data recorded in coincidence with Br $K\alpha$ and $K\beta$ x-rays, we estimate the uncertainties in the measured energies to be $\sim$10\%. 

Bromine has two approximately equally abundant isotopes, $^{79}$Br and $^{81}$Br, while iodine has the single isotope, $^{127}$I. In the present work, we are interested in replacing one of the Br atoms of Br$_2$ with the much heavier I atom. In this case, we report calculations and measurements for $^{79}$Br/$^{79}$Br and $^{79}$Br/$^{127}$I only.

\subsection{THEORY\label{theory}}
We model x-ray–induced inner-shell cascades with Monte Carlo (MC)/Molecular dynamics (MD) \cite{Ho2017:jpb, Ho2020:natcomm} coupled with a classical over-the-barrier (COB)\cite{Ryufuku1980PRA,PhysRevLett.113.073001,Ho23:jcp, Boll2022} framework. 
The MC module samples inner-shell Auger–Meitner cascades and valence-hole dynamics across charge states using relativistic Hartree-Fock-Slater (HFS) electronic structure, from which we compute photoionization cross sections, shake-off, and decay rates for Auger–Meitner and fluorescence. 
Shake-off\cite{SVENSSON1988327, kochur1995direct} accounts for an additional ionization of a valence electron due to a large perturbation in the valence region after photoionization or Auger decay. As will be shown in Table~\ref{tab}, shake-off is responsible for creating high-charge product channels (e.g., total charge of 5 for $K\beta$ and total charge of 8 for $K\alpha$) that cannot be reached without it.
The MD module tracks atomic, ionic, and electron motion, while COB handles interatomic valence electron transfer. 
The COB model has previously been applied to molecular fragmentation dynamics in intense EUV\cite{PhysRevLett.113.073001} and x-ray fields\cite{Rudenko2017,doi:10.1126/science.1253607,Boll2022}.
This coupled MC/MD-COB approach enables tracking of inner-shell Auger-Meitner cascades, electron transfers in valence orbitals, and the fragmentation of molecules across multiple timescales \cite{Ho23:jcp}.

In the original COB model, electron transfer between electron donor ($D$) and acceptor $A$ pair separated by a distance $R_{DA}$ at time $t$ is assumed to occur instantaneously when the electron binding energy, $BE_D$, is higher than the Coulomb barrier, $V_{b}(R_{DA})$, of the D-A pair. We replace the instantaneous COB criterion with a probabilistic transfer applied at each time step $\delta t$ \cite{XLi2025}, 
\begin{equation}
    P_{ET} = \delta t / t_{ET},
\label{Eq1}
\end{equation}
where $t_{ET}$ is the time required for an electron with kinetic energy, $BE_D - V_{b}(R_{DA}$), to traverse the separation $R_{DA}$ (see Fig. \ref{sch1}).  
Accordingly, the binary (0 or 1) probability of electron transfer in the original scheme, is adjusted by the binding energetics (energy-level offset) and the increasing donor–acceptor separation during dissociation.
We use $\delta t =$ 10 as for the first 100 fs and 1 fs there after. 
Note that, in our model, 
the partner Br or I atom is charged only by electron transfer occurring between the outermost valence shells—Br 4p and I 5p.

We simulate Br$_2$ and IBr with a single Br 2p ($K\alpha$) or 3p ($K\beta$) vacancy created after K-shell fluorescence, using 10$^{5}$ trajectories per case. For $K\beta$ cases that end with a total charge of 5  (Br$^{2+}$Br$^{3+}$ or I$^{2+}$Br$^{3+}$), we implement our simulation from [3d$^{-2}$] electron configuration because our HFS model predicts that the 3p$^{-1} \to$ 3d$^{-2}$ Auger step is energetically forbidden.
For MC/MD-COB simulation starting with 3p$^{-1}$ configuration, only 5 to 6 of 10${^5}$ trajectories reach the total‑charge‑5 channel, only via two Auger decays and two shake‑off events. This is a physically unfavorable pathway and its very low yield makes analysis of this channel computationally intractable.
However, our supplementary HF/6-311G(d,p) calculations on Br$_2$ and IBr indicate that the 3p$^{-1} \to$ 3d$^{-2}$ transition is energetically allowed; accordingly, we analyze 5 $\times$ 10$^{4}$ trajectories for these Br$^{2+}$Br$^{3+}$ and I$^{2+}$Br$^{3+}$ channels, starting with 3d$^{-2}$ configuration. Detailed descriptions of the computational method are available in the Supporting Information.

\section{RESULTS AND DISCUSSION\label{res}}
Figure~\ref{sch1} shows a schematic of core-hole relaxation and molecular fragmentation in Br$_2$ and IBr: (i) the neutral molecule is photoionized by synchrotron X-rays, ejecting a Br 1s electron; (ii) the core hole is filled via X-ray fluorescence, either $K\alpha$ ($2p \to 1s$) or $K\beta$ ($3p \to 1s$); (iii) subsequent relaxation proceeds through successive Auger decays that emit electrons, and electron transfer between atomic sites can ionize the partner Br or I atom; and (iv) the resulting positive charges drive molecular fragmentation through Coulomb repulsion.

\begin{figure*}[h]
\includegraphics[width=14cm]{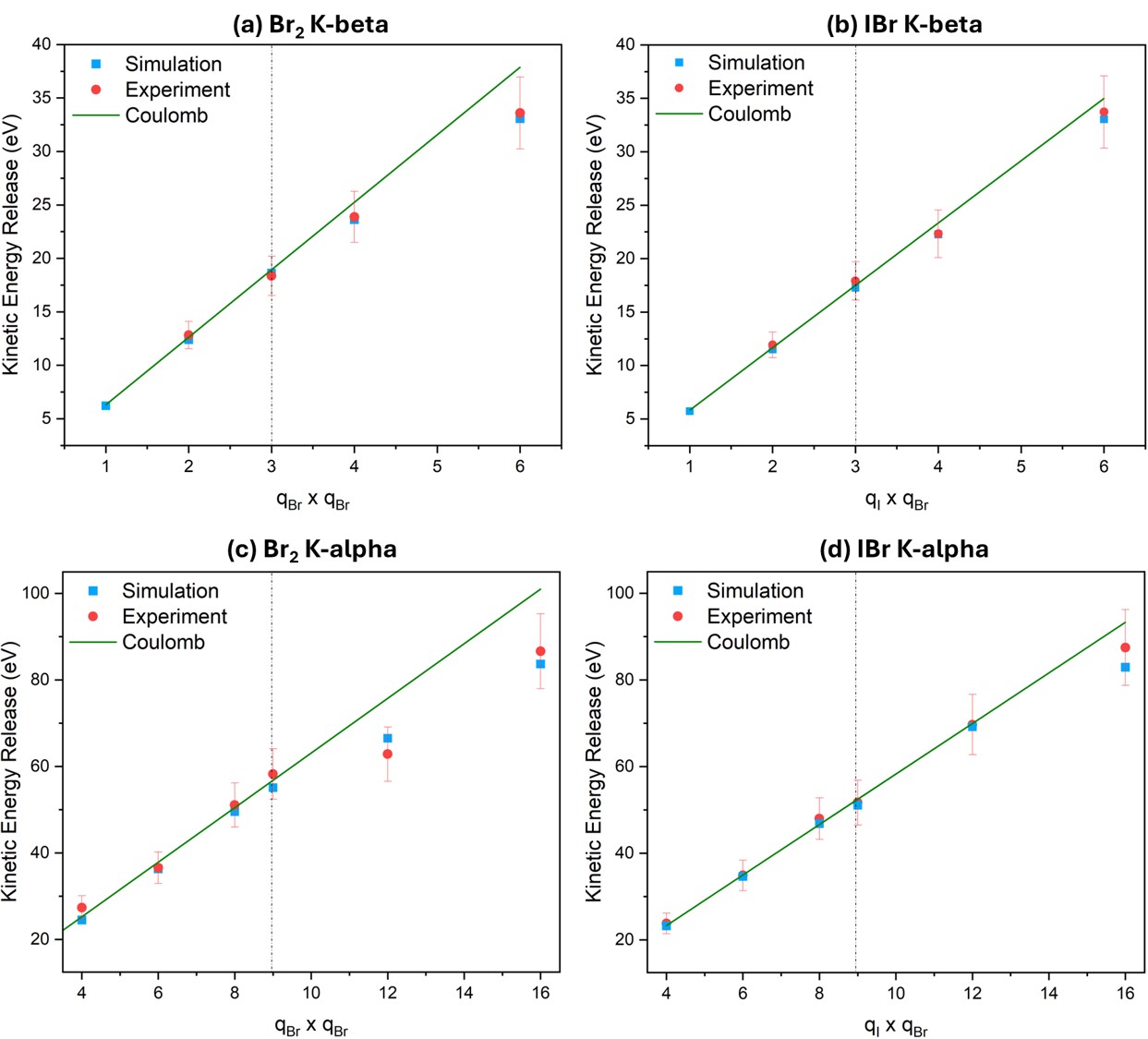}
\caption{Kinetic energy releases of various ion fragmentation pairs of Br$_2$ (a,c left panels) and IBr (b,d right panels) measured in coincidence with Br K$\beta$  (a,b, top panels) and K$\alpha$ (c,d bottom panels) at 13,486 eV emission. The green lines represent the calculated Coulomb energy for the respective charge-state products, based on the equilibrium internuclear distances: R = 2.281 \r{A} (Br$_2$) and 2.469 \r{A} (IBr). 
The red circles denote experimental data, while the blue squares indicate calculated kinetic energy release. The red error bar indicates the 10 percent uncertainty of experimental measurements. The measured and calculated results are for $^{79}$Br: the results for $^{81}$Br are similar but not plotted. }
\label{Fig2}
\end{figure*}

Figure~\ref{Fig2} presents the kinetic energy release (KER) distributions for Br$_2$ (left panels: a and c) and IBr (right panels: b and d), following Br 1s ionization and coincident detection of $K\beta$ (top panels: a and b) or $K\alpha$ (bottom panels: c and d) X-ray emission. The KER is plotted as a function of the charge product, $q_{\text{Br/I}} \times q_{\text{Br}}$, where $q_{\text{Br}}$ and $q_{\text{I}}$ denote the final charge states of Br and I, respectively, labeled as ($q_{\text{Br/I}}$, $q_{\text{Br}}$). For simplicity, only the $^{79}$Br isotope is considered, as $^{81}$Br yields similar trends with slightly reduced KER due to its higher atomic mass. 

Overall, the simulated results (blue squares) show good agreement with experimental data (red circles). For comparison, Coulomb energies (green lines) are also plotted, computed using equilibrium internuclear distances of 2.281~\r{A} for Br$_2$ and 2.469~\r{A} for IBr. 
This shows that our semi-classical MC/MD-COB simulation reasonably demonstrates its accuracy and versatility in modeling inner-shell cascade processes.

We observe that KER generally follows Coulomb energy up to a charge product of 3 for $K\beta$ and 9 for $K\alpha$ (indicated by vertical dotted lines), where the final KER closely matches the Coulomb energy (C.E.), suggesting that the final charge state is reached with minimal nuclear motion. However, at higher charge products, the KER deviates from the C.E.—evident in the (3,2) channel for $K\beta$ (Fig. \ref{Fig2}a and b) and (4,3), (4,4) channels for $K\alpha$ (Fig. \ref{Fig2}c and d), indicating significant nuclear motion. This deviation is more pronounced in Br$_2$ than in IBr.

These differences in KER can be attributed to two main factors: (1) \textit{electronic effects}, referring to differences in inner-shell cascade dynamics between Br$_2$ and IBr, particularly the competition between electron transfer and Auger decay; and (2) \textit{nuclear effects}, referring to differences in atomic mass, where the heavier iodine atom dissociates more slowly than bromine.

\subsection{Electronic Effects}
To investigate electronic effects in greater detail, we examined the dominant inner-shell cascade pathways, as summarized in Table~1. Specifically, to understand the KER deviation from C.E. in the (3,2) $K\beta$ and (4,4) $K\alpha$ channels, we analyzed the Br electron configuration during the cascade initiated by a 1s core hole, followed by either $K\beta$ or $K\alpha$ emission (see Figure~\ref{Fig3}).

We find that the dominant cascade sequence and pathways are the same between Br$_2$ and IBr, regardless of the final charge product. Figure~\ref{Fig3} also shows that both Auger decay timing and the associated electron configurations of Br are comparable between the two molecules, suggesting that differences in electronic effect does not critically influence KER. 

Figure~\ref{Fig3} shows that, in all cases, deeper core holes such as 2p, 3s, and 3p decay within sub-femtosecond timescales, while 3d holes are longer-lived, with lifetimes ranging from several femtoseconds to over 10~fs, allowing nuclei to dissociate. The importance of these 3d holes has been examined in our previous study comparing inner-shell cascade dynamics of IBr upon 1s hole creation at either the I or Br site \cite{Ho23:jcp}.

\subsection{Nuclear Effects}
\begin{figure*}[h]
\includegraphics[width=16cm]{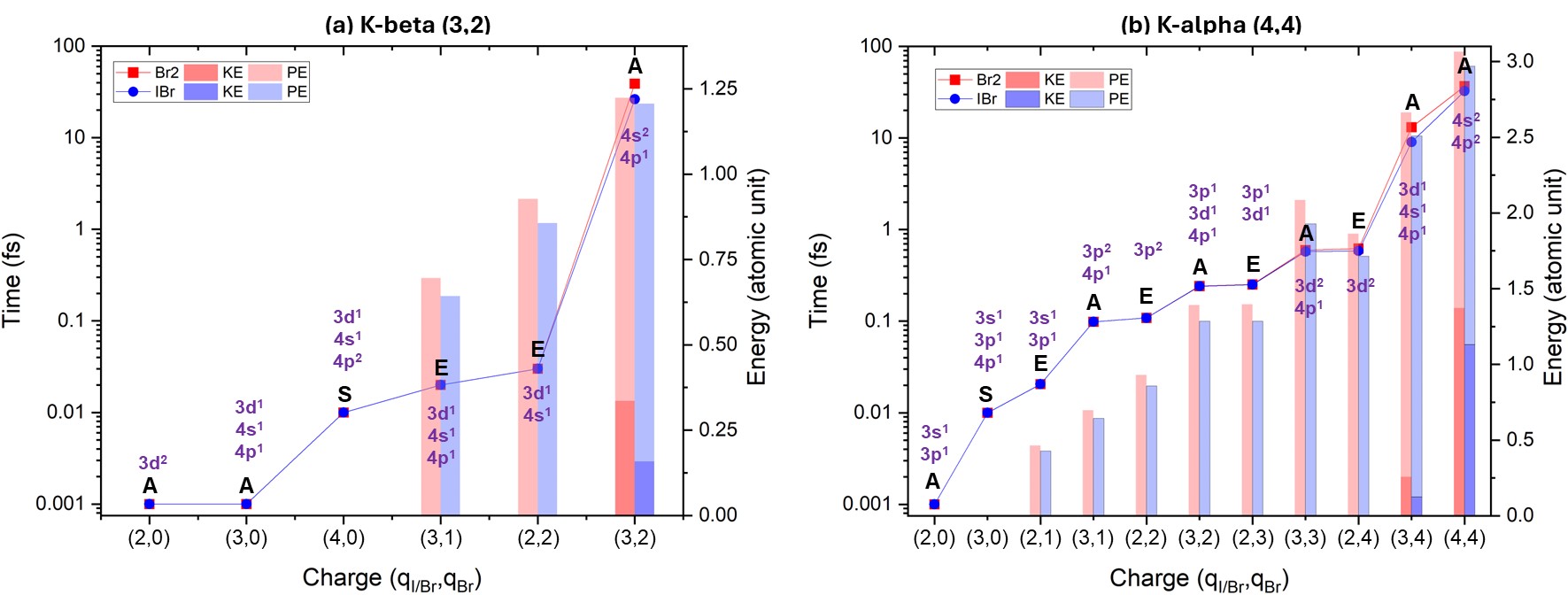}
\caption{Sequence and timing of inner-shell decay (curves, left y-axis) for (a) the K$\beta$ (3,2) channel and (b) the K$\alpha$ (4,4) channel for Br$_2$ (red and pink) and IBr (blue and skyblue).
A, S, and E denote Auger decay, shake-off, and electron transfer, respectively. The resulting hole configurations of the Br atom, which underwent 1s core-ionization followed by K$\beta$/K$\alpha$ emission, are shown. The stacked bar graphs (right y-axis) indicate the average potential and kinetic energy associated with each charge distribution at the moment that distribution is reached.
This figure highlights the ultrafast lifetimes of inner-shell (3s, 3p) core holes and the relatively long lifetime of 3d holes, which allows for conversion of potential energy into kinetic energy.
}
\label{Fig3}
\end{figure*}

In Figure~\ref{Fig3}, the evolution of kinetic and potential energies of $K\beta$ (3,2) channel (panel a) and $K\alpha$ (4,4) channel (panel b) during the cascade is illustrated, providing insight into nuclear effects on KER. 
Once both atoms are ionized, Coulombic potential energy is stored and gradually converted into kinetic energy as the nuclei dissociate.

This energy conversion is particularly evident during Auger decay from 3d holes—for example, during the (2,2) $\rightarrow$ (3,2) transition in $K\beta$, and (2,4) $\rightarrow$ (3,4) $\rightarrow$ (4,4) transitions in $K\alpha$. Deeper core holes decay too rapidly for nuclear motion to occur, whereas the relatively long-lived 3d holes permit significant dissociation during their lifetime. 

Here, the combination of stronger Coulomb forces and the lower mass of Br plays a significant role.
Coulomb potential energy is inversely proportional to the internuclear distance, and the shorter equilibrium bond length of Br$_2$ (2.281~\r{A}) compared to IBr (2.469~\r{A}) leads to stronger repulsion in Br$_2$ for a given charge product. Furthermore, the velocity of dissociation is inversely proportional to the square root of atomic mass, so Br$_2$ dissociates approximately 11\% faster than IBr under the same Coulombic force.

As a result, during the lifetime of the 3d$^{-1}$ core hole, Br$_2$ dissociates from 2.52~\r{A} to 3.46~\r{A}, whereas IBr dissociates only from 2.49~\r{A} to 2.58~\r{A}. This effect is especially pronounced in the $K\alpha$ (4,4) channel, where 3d$^{-2}$ holes enable more extensive Br$_2$ bond breaking. In the $K\beta$ (3,2) channel, the 3d$^{-2}$ hole arises directly from a 3p$^{-1}$ hole, but since the partner atom (Br or I) is not ionized, the transition from 3d$^{-2}$ to 3d$^{-1}$ does not contribute to nuclear dissociation. The same argument applies to the (4,3) channel in $K\alpha$.

\begin{table*}[h]
\centering
\begin{tabular}{ |c|c|c|c|c|c|c|c|c| } 
 \hline
  & \multicolumn{4}{| c |}{Br$_2$ (K$\beta$)} & \multicolumn{4}{| c |}{IBr (K$\beta$)} \\
 \hline
  ($q_{I/Br}$,$q_{Br}$) & KER (Exp) & KER(Theo) & C.E. &  Seq & KER (Exp) & KER(Theo) & C.E. & Seq \\
 \hline
 \hline
 (1,1) & - & 6.2 & 6.313 & AE & - & 5.7 & 5.8 & AE \\
 (1,2) & 12.8 & 12.4 & 12.6 & AEA & 11.9 & 11.5 & 11.7 & AEA \\
 (1,3) & 18.4 & 18.6 & 18.9 & AAEA & 17.9 & 17.3 & 17.5 & AAEA \\
 (2,2) & 23.9 & 23.6 & 25.3 & AAEAE & 22.3 & 22.3 & 23.3 & AAEAE \\
 (2,3) & 31.4 & 33.1 & 37.9 & AASEEA & 33.7 & 33.0 & 35.0 & AASEEA \\
 \hline
 \hline
 & \multicolumn{4}{| c |}{Br$_2$ (K$\alpha$)} & \multicolumn{4}{| c |}{IBr (K$\alpha$)} \\
 \hline
  ($q_{I/Br}$,$q_{Br}$) & KER (Exp) & KER(Theo) & C.E. &  Seq & KER (Exp) & KER(Theo) & C.E. &  Seq \\ 
 \hline
 \hline  
 (2,2) & 27.4 & 24.5 & 25.3  & AAEAE       & 23.8 & 23.2 & 23.3 & AAEAE       \\
 (2,3) & 36.6 & 36.3 & 37.9  & AAEAEA      & 34.9 & 34.6 & 35.0 & AAEAEA      \\
 (2,4) & 51.1 & 49.6 & 50.5  & AAEAEAA     & 48.0 & 46.8 & 46.7 & AAEAEAA     \\
 (3,3) & 58.3 & 55.1 & 56.8  & AAEAEAEA    & 51.7 & 51.1 & 52.5 & AAEAEAEA    \\
 (3,4) & 62.9 & 66.5 & 75.8  & ASEAEAEAA   & 69.7 & 69.2 & 70.0 & ASEAEAEAA   \\
 (4,4) & 86.6 & 83.7 & 101.0 & ASEAEAEAEAA & 87.5 & 82.9 & 93.3 & ASEAEAEAEAA \\
 \hline
 \end{tabular}
\caption{Kinetic energy release (KER) from experiment (Exp) and theory (Theo), along with Coulomb explosion energy (C.E.) calculated based on the equilibrium bond lengths of Br$_2$ (2.281~\r{A}) and IBr (2.469~\r{A}). Also shown is the sequence (Seq) of inner-shell decay processes, where A, S, and E denote Auger decay, shake-off, and electron transfer, respectively. Both KER and C.E. values are given in eV.}
\label{tab}
\end{table*}

\begin{figure*}[h]
\includegraphics[width=16cm]{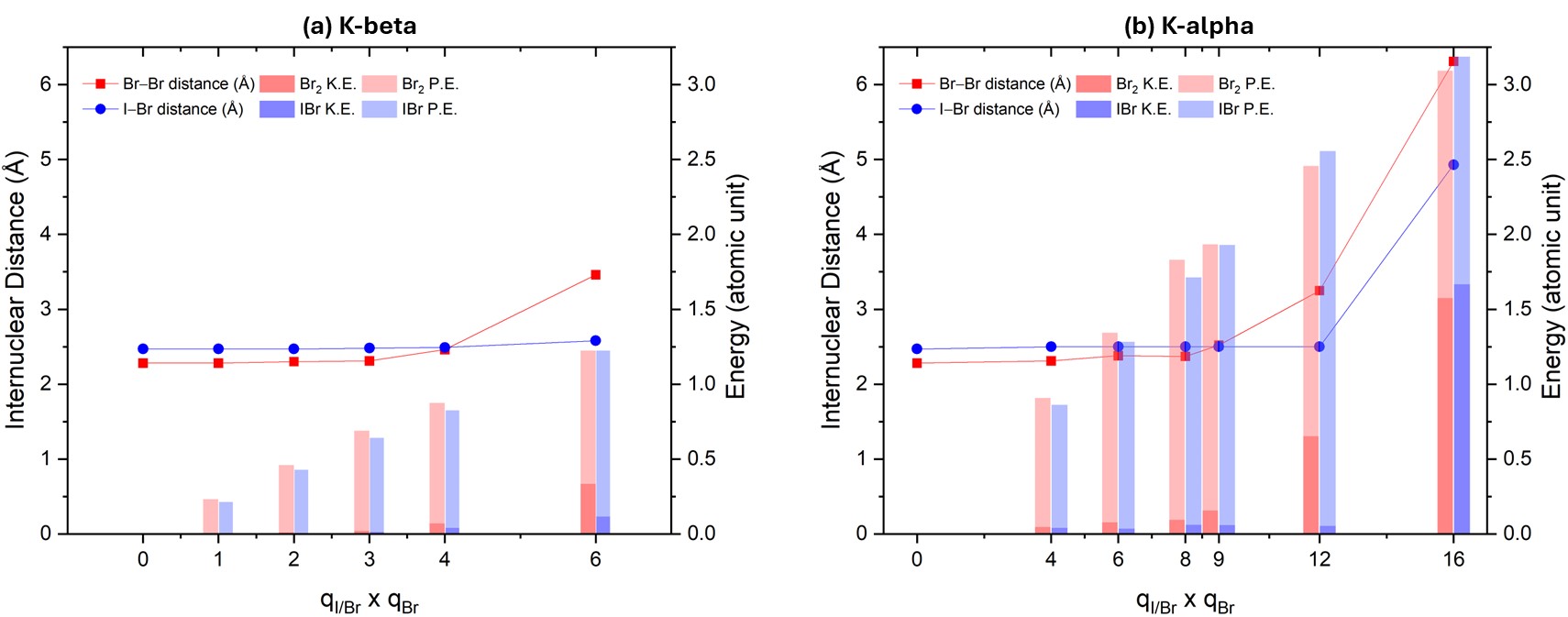}
\caption{Internuclear distance (curves, left y-axis) and potential/kinetic energy distribution (stacked bar graphs, right y-axis) of Br$_2$ (red and pink) and IBr (blue and skyblue) at the moment the charge distribution (charge product) $q_{\text{I/Br}} \times q_{\text{Br}}$ is created.  
This figure demonstrates the deviation in KER observed in Fig.~\ref{Fig2} as a function of increasing charge product. The deviation arises from the conversion of potential to kinetic energy that occurs during the long-lived 3d-hole states, allowing nuclear motion prior to complete charge redistribution.}
\label{Fig4}
\end{figure*}

\section{CONCLUSIONS\label{con}}
In this work, we studied the inner-shell decay and fragment dynamics of Br$_2$ and IBr molecules following Br 1s ionization, focusing on the interplay between electronic and nuclear effects. 
Our experimental and theoretical results demonstrate that the KER of fragment ions deviates from Coulomb explosion predictions, particularly at higher charge states. These deviations arise from the sequential nature of inner-shell cascades and the coupling between electronic transitions and nuclear motion. 

The comparison between Br$_2$ and IBr highlights the significant role of atomic substitution, where the heavier iodine atom in IBr leads to slower nuclear motion, resulting in a KER more comparable to a Coulomb explosion than in Br$_2$.
This difference arises mainly from the atomic mass effect, where the velocity of dissociation is inversely proportional to the square root of atomic mass. 
This is particularly pronounced during the lifetime of long-lived core-hole states, such as the 3d hole in Br, which allows for substantial nuclear motion in Br$_2$
but only limited motion in IBr.

The significance of electron–nuclear coupling is well recognized in light systems, where the difference between H and D manifests as a kinetic isotope effect. However, it has been less intuitive that similar effects also play a critical role in heavy elements. In this sense, our findings underscore the importance of considering both electronic and nuclear effects in the interpretation of molecular fragmentation dynamics in heavy-element systems.
Charge redistribution from decay of inner-shell vacancy states in heavy atoms of polyatomic molecules may provide further insight into electronic and nuclear motion processes.
The atomic mass effect, e.g. isotope effect, can play a critical role in determining the extent of nuclear motion during inner-shell cascades, influencing the KER distributions. 
Future studies could extend this approach to other heavy-element systems and investigate the signatures of isotope substitution on KER. As demonstrated by xenon isotopes in nuclear test monitoring, isotope-specific effects can provide valuable insights into underlying nuclear and electronic dynamics \cite{bowyer2002detection,le2013innovative}.

\section{SUPPLEMENTARY MATERIAL}
The Supplementary Information details the experimental procedures and simulation methods.

\begin{acknowledgments}
This work was supported by the U.S. Department of Energy, Office of Science, Basic Energy Sciences, Chemical Sciences, Geosciences, and Biosciences Division. Use of the Advanced Photon Source, an Office of Science User Facility operated for the U.S. Department of Energy (DOE) Office of Science by Argonne National Laboratory, was supported by the U.S. DOE under Contract No. DE-AC02-06CH11357. We gratefully acknowledge the computing resources provided on Improv, a high-performance computing cluster operated by the Laboratory Computing Resource Center at Argonne National Laboratory.
\end{acknowledgments}

\section{Author Declaration}
\subsection{Conflict of Interest}
The authors have no conflicts to disclose.
\subsection{Author Contributions}
\textbf{Nivedita Bhat}: Formal Analysis (equal), Investigation (equal), Software (equal), Validation (equal), Writing – original draft (equal)
\textbf{Yeonsig Nam}: Conceptualization (equal), Methodology (equal), Formal Analysis (equal), Investigation (equal), Software (equal), Validation (equal), Writing – original draft (equal), Supervision (equal)
\textbf{Linda Young}: Conceptualization (equal), Funding Acquisition, Project Administration, Writing – original draft (equal)
\textbf{Stephen H. Southworth}: Conceptualization (equal), Methodology (equal), Formal Analysis (equal), Investigation (equal), Validation (equal), Writing – original draft (equal), Supervision (equal)
\textbf{Phay J. Ho}: Conceptualization (equal), Methodology (equal), Investigation (equal), Software (equal), Writing – original draft (equal), Supervision (equal)

\section{Data Availability Statement}
The data that support the findings of this study are available from the
corresponding author upon reasonable request.

\section{References}
\bibliography{IBr_Br2_JCP}

\end{document}